\documentclass[aps,showpacs]{revtex4}
\usepackage{amssymb}
\usepackage[dvips]{graphicx}
\usepackage[english]{babel}
\usepackage{indentfirst}
\usepackage{amsxtra}
\usepackage{amsmath}
\usepackage{supertabular}
\usepackage{multirow}
\usepackage [mathcal]{eucal}
\newcommand {\ortt}{\omega^{\rm RPA}_{\rm tt} }
\newcommand {\ortn}{\omega^{\rm RPA}_{\rm tn} }
\newcommand {\ornt}{\omega^{\rm RPA}_{\rm nt} }
\newcommand {\ornn}{\omega^{\rm RPA}_{\rm nn} }
\newcommand {\oit}{\omega^{\rm IPM}_{\rm t}}
\newcommand {\oin}{\omega^{\rm IPM}_{\rm n}}
\newcommand {\car}{$^{12}$C}
\newcommand {\oxy}{$^{16}$O}
\newcommand {\caI}{$^{40}$Ca}
\newcommand {\caII}{$^{48}$Ca}
\newcommand {\zir}{$^{90}$Zr}
\newcommand {\lead}{$^{208}$Pb}
\def\beq{\begin{equation}}
\def\eeq{\end{equation}}
\def\beqn{ \begin{eqnarray} }
\def\eeqn{ \end{eqnarray} }
\def\s1s2{{ \boldsymbol{\sigma}(1) \cdot \boldsymbol{\sigma}(2) }}
\def\t1t2{{ \boldsymbol{\tau}(1) \cdot \boldsymbol{\tau}(2)  }}
\newcommand{\bsigma}{\mbox{\boldmath $\sigma$}}

\newcommand{\br}{{\bf r}}
\newcommand{\bq}{{\bf q}}

\begin{document}
\noindent
\title{Tensor effective interaction in 
 self-consistent Random Phase Approximation calculations}
\author{M. Anguiano$^1$, G. Co'$\,^{2,3}$, V. De Donno$\,^{2,3}$ 
and  A. M. Lallena$^1$}
\affiliation{
\mbox {1) Departamento de F\'\i sica At\'omica, Molecular y
  Nuclear,} \\ 
\mbox {Universidad de Granada, E-18071 Granada, SPAIN } \\
 \mbox {2) Dipartimento di Fisica, Universit\`a del Sa lento,
Via Arnesano, I-73100 Lecce, ITALY} \\
\mbox {3) INFN, Sezione di Lecce, Via Arnesano, I-73100 Lecce, ITALY }
}
\date{\today}

\bigskip

\begin{abstract} 
We present a 
study of the effects of the tensor-isospin term of the effective
interaction in Hartree-Fock and Random Phase Approximation
calculations. We used finite-range forces of Gogny type, and we
added to them a tensor-isospin term which behaves, at large
internucleonic distances, as the analogous term of the microscopic
interactions.  The strength of this tensor force has been chosen to
reproduce the experimental energy of the lowest 0$^-$ excited state in
$^{16}$O, which shows large sensitivity to this term of the
interaction.  With these finite-range interactions, we have studied
the effects of the tensor-isospin force in ground and excited states
of carbon, oxygen, calcium, nickel, zirconium, tin and lead
isotopes. Our results show that the tensor force affects mainly the
nucleon single particle energies. However, we found some interesting
cases where also bulk nuclear properties are sensitive to the tensor
interaction.
\end{abstract}

\bigskip
\bigskip
\bigskip

\maketitle

\section{Introduction}
\label{sec:intro} 
At the beginning of the 40's of the past century, the existence of the
electric quadrupole moment of the deuteron \cite{kel39,nor40} was
explained by Rarita and Schwinger by introducing a static tensor term
in the nucleon-nucleon (N-N) interaction \cite{rar41a,rar41b}. Since
then, tensor terms are unavoidable ingredients of the microscopic N-N
interactions, i.e. those interactions constructed to reproduce the
properties of two-nucleon systems.

Despite their relevance in microscopic interactions, the tensor terms
are usually neglected when effective interactions and theories are
used. In the effective theories some complicated many-body effects are
treated, obviously in effective and average manner, by changing the
values of the parameters of the interaction. Some specific observables
of the nucleus are chosen to select these values. For example, in
Hartree-Fock calculations these observables are usually the nuclear
binding energies. The effective theory is expected to be able to
describe other observables. If this fails, one searches for many-body
effects which should be explicitly treated to improve the description
of the data. In this manner, we link many-body effects to specific
observables. In the case of our interest here, the tensor force, the
point is the identification of observables clearly depending on the
presence of this term in the effective N-N interaction.

In these last years, the interest on the tensor terms of the effective
N-N interaction has increased because the inclusion of these terms
improves the description of the single particle (s.p.) energies of
some isotope or isotone chains \cite{oza00,kan02,sch04,sor08} when the
Hartree-Fock theory is used
\cite{ots05,ots06,bri07,sug07,tar08,col07,cao09}. 

We see some weak points in using the s.p. energies to define the
strength of the tensor terms of the effective interactions. First,
s.p. energies are extremely sensitive to the spin-orbit terms of
the N-N interaction, and this obscures the possibility of a clear
identification of the tensor effects (see for example the discussion
in Ref. \cite{les07}). Furthermore, observations are always done on
global nuclear properties, therefore the identification of the
measured quantities with s.p. properties of the nucleus is done by
imposing to the observed quantity the physical interpretation given
within a mean field description of the many-body system.  The fact
that experimental values of spectroscopic factors are usually rather
different from the mean field expectations is a clear indication of
the limits of this procedure.

In this article we propose an alternative approach to select the
strength of the effective tensor forces. We have looked at the
excitation spectrum to find observables particularly sensitive to the
tensor force. We have identified these observables with the energies
of the 0$^-$ charge conserving excitations. 
The large sensitivity of
the 0$^-$ excitation to the tensor parts of the nucleon-nucleon
interaction was pointed out already in Ref. \cite{blo68}.  By using a
recursive self-consistent Hartree-Fock (HF) plus Random Phase
Approximation (RPA) procedure, we have chosen the strength of the
tensor term of the effective interaction to reproduce the experimental
value of the 0$^-$ in $^{16}$O. With these new interactions we have
investigated the ground and excited state properties of various nuclei
by doing HF plus RPA calculations.

The structure of the tensor terms we have considered and the
methodology used to select their strengths are presented in
Sec. \ref{sec:int}. We discuss the results obtained in the
description of the ground states of various nuclei in Sec.
\ref{sec:gs}, and in Sec. \ref{sec:rpa} the results obtained for the
excited states. We summarize the main points of our work and draw our
conclusions in Sec. \ref{sec:conc}.

\section{The interaction}
\label{sec:int} 

The most important tensor component of the microscopic N-N interaction
is that related to the tensor-isospin channel \cite{mac87,wir95} whose
long range behaviour is dominated by the exchange of a single
pion. Since the pion is the lightest meson, the range of the
tensor-isospin term is longer than the ranges of the other terms of
the N-N interaction.  For this reason we have chosen to consider, in
our effective interactions, only tensor-isospin terms with finite
range.  The use of finite range forces requires, in both HF and RPA
calculations, the evaluation of direct and exchange interaction matrix
elements.

\begin{figure}[t]
\begin{center}
\includegraphics[scale=0.4]{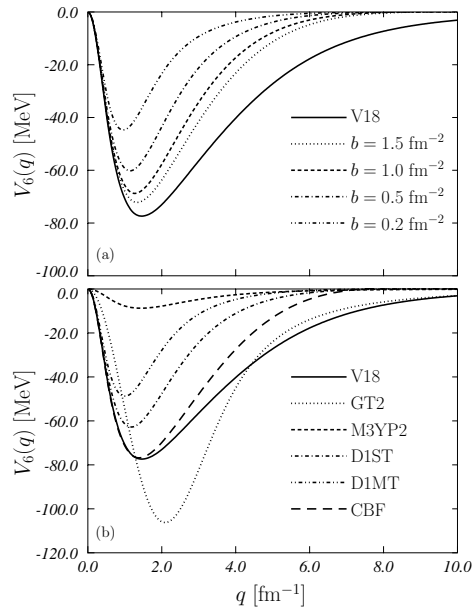} 
\caption{\small Momentum dependent term of the tensor force, see 
Eq. (\ref{eq:fourier}), for various parameterizations. 
In the panel (a) we compare the tensor-isospin
term of the bare Argonne V18 interaction (full line) with the
interactions obtained from Eq. (\ref{eq:vtensor}) by using different
values of the parameter $b$.  In panel (b) we compare the bare Argonne
V18 term with the tensor term of the GT2 force of Ref. \cite{ots06}
(dotted line) and with that of the M3YP2 force of Ref. \cite{nak03}
(short-dashed line). The tensor forces we have constructed, named
D1ST and D1MT, are indicated by the dashed-dotted and
dashed-doubly-dotted lines. The long-dashed line (CBF) shows the
tensor term obtained by multiplying the bare Argonne V18 interaction
with the scalar part of the 
correlation function obtained in microscopic Correlated Basis
Function calculations \cite{ari07}.  }
\label{fig:tensq}
\end{center}
\end{figure}

The tensor-isospin term of our effective interaction is based on the
analogous term of the microscopic Argonne V18 interaction
\cite{wir95}.  We have multiplied the radial part of this term by a
function which simulates the effect of the short-range correlations
\cite{ari07}. In our work, the radial part of the tensor-isospin term
has the form
\beq
v_6(r) \,= \, v_{6,{\rm AV18}}(r) \, \left[ 1\, -\, 
\exp \left(-\,b \, r^2 \right) \right] \, ,
\label{eq:vtensor}
\eeq
where we have indicated with $r$ the distance between the two
interacting nucleons, with $ v_{6,{\rm AV18}}$ the radial function of
the Argonne V18 tensor-isospin potential \cite{wir95}, and with $b$ a
free parameter.  The changes in the tensor-isospin term produced by
choosing different values of $b$ are shown in  panel (a) of
Fig. \ref{fig:tensq}, where we present the Fourier transformed
function $V_6(q)$, defined by the equation
\begin{eqnarray}
V_6(q)\,S_{12}(\bq) &= & \nonumber
\int {\rm d}^3 r \, \exp \left( i\, \bq\cdot\br \right) \, 
v_6(r)\, S_{12}(\br) \\ &
=&  - \, 4 \pi \int {\rm d}r \, r^2 \, j_2(qr) \, v_6(r) \, S_{12}(\br)  
\, .
\label{eq:fourier}
\end{eqnarray}
In the above equation $r$ and $q$ indicate the moduli of $\br$ and
$\bq$, and we used the definition of the tensor operator
\beq
S_{12}(\br) \, = \, 
3 \, \frac{[\bsigma(1)\cdot\br] \,[\bsigma(2)\cdot\br]}{r^2} \,
- \, \bsigma(1)\cdot\bsigma(2)
\, ,
\label{eq:tensor}
\eeq
where $\bsigma$ are the usual Pauli spin matrices.

The results presented in the panel (a) of Fig. \ref{fig:tensq} show
that the correlation effects reduce the strength of the bare tensor
force. The smaller is the value of $b$, the more extended in $r$ space
is this effect, therefore the function to be integrated in
Eq. (\ref{eq:fourier}) becomes smaller.

The first step of our study consisted in identifying an observable
very sensitive to the tensor force with the aim of 
using it to determine the
strength of this part of the effective interaction.  We have conducted
this study by investigating the excitation spectra of the \car, \oxy,
\caI, \caII, \zir \/ and \lead \/ nuclei within the phenomenological
RPA approach developed by the J\"ulich group \cite{spe77,spe80}. In
this approach, based on the Landau-Migdal theory of finite Fermi
systems \cite{mig67}, the set of s.p. energies and states is obtained
by using phenomenological mean field potentials that, in our case,
have the shape of Woods-Saxon wells. The values of the parameters of
the potential are chosen for each nucleus in order to reproduce at
best the empirical values of the s.p. energies around the Fermi
surface and the values of the charge root mean square radii. The
explicit expression of the potential, and the values of the
parameters, can be found in Refs. \cite{ari07,co09b}. Following the
philosophy of the Landau-Migdal approach, in the RPA calculations we
substituted the Woods-Saxon s.p. energies with their experimental
values, when they are available.

All the RPA results presented in this article have been obtained by
using a discrete s.p. basis. In the phenomenological calculations
discussed in this section we have used a discrete s.p. basis obtained
by diagonalizing the Woods-Saxon well in a harmonic oscillator
basis. In analogy with the work of Ref. \cite{co09b}, we have used
configuration spaces large enough that the inclusion of additional
s.p. states does not modify the energies of the first excited states,
below 20 MeV in the lighter nuclei and 15 MeV in the heavier ones,
within 0.1 MeV.

In these phenomenological calculations we used as basic N-N effective
interaction a density dependent Landau-Migdal force. The values of the
force parameters, given in Refs. \cite{co09b,don09}, change for each
nucleus, and are chosen to reproduce at best the energies of the
collective low-lying 3$^-$ states of each nuclei, and also the
energies of the 12$^-$ excited states in $^{208}$Pb. We added to this
basic interaction a tensor term of the form given by
Eq. (\ref{eq:vtensor}), and we studied the excitation spectra of the
nuclei mentioned above.

In agreement with previous calculations \cite{spe80,co90} we observed
that the influence of the tensor term on the natural parity states is
negligible. We found few cases of interest in the spectrum of the
unnatural parity excitations. The most interesting one was the
excitation of the 0$^-$ states, which show common characteristics in
all the nuclei we considered.

\begin{figure}
\begin{center}
\includegraphics[scale=0.4]{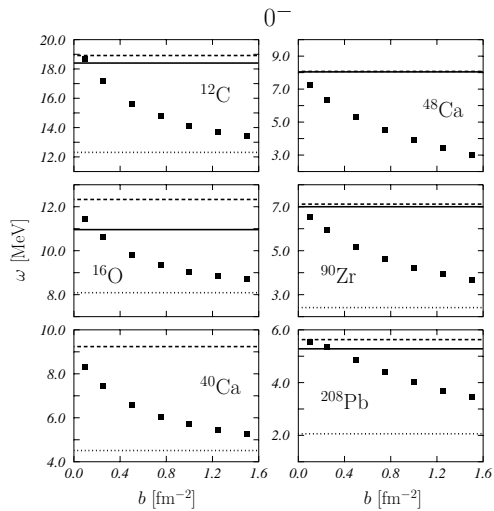} 
\caption{\small The squares show the values 
 of the excitation energies of the lowest 0$^-$
 states for various nuclei 
 as a function of the value of the parameter $b$ of Eq.
 (\ref{eq:vtensor}) ruling the strength of the tensor force. The full
 lines show the experimental values, the dashed lines the values
 obtained without tensor, and the dotted lines the values obtained
 with the full tensor term of the Argonne V18 interaction. 
 }
\label{fig:ene0-}
\end{center}
\end{figure}

In Fig. \ref{fig:ene0-} we have summarized the results obtained for
the energies of the lowest 0$^-$ excitations in the nuclei
investigated. In this figure, the squares indicate the values of the
$0^-$ excitation energies, obtained by adding the tensor term of
Eq. (\ref{eq:vtensor}) to the Landau-Migdal force, as a function of
the parameter $b$. The horizontal dashed lines indicate the values
obtained without tensor term, while the dotted lines show the values
obtained by using the full tensor term of the Argonne V18 interaction.
The full lines show the experimental values \cite{bnlw,heu07}.

We observe that the effect of the tensor term is always attractive,
i.e.  all the energies obtained with the tensor terms are smaller
than those obtained without it. The values of these energies decrease,
monotonically and smoothly, from the dashed to the dotted lines when
the value of $b$ increases, as we have naively expected. Discussing the
results in more detail, we have obtained variations of the energies
from 4 up to 8 MeV, and, in relative variations, between 30\% and
100\%. These are the largest effects of the tensor force on the
excitation energies we have found in our investigation.  In the \caII
\/ case, the effect of the full tensor term is so strong that we
obtained an imaginary solution for the RPA equations.

The remarkable sensitivity of the $0^-$ excitation energies to the
presence of the tensor force, and their smooth behaviour with respect
to the changes of its strength, make these energy values particularly
suitable to be chosen as experimental benchmarks to select the
strength of the tensor terms in effective interactions.

In this work we have performed HF plus RPA calculations. The HF
equations were solved with the method used in
Refs. \cite{bau99,co98b}. This method is based on the plane wave
expansion technique developed by Guardiola and Ross
\cite{gua82a,gua82b}. After the iterative process has reached the
convergence, we solved the HF differential equations not only for the
s.p. states below the Fermi level, the hole states, but also for those
states above the Fermi level, the particle states.  The numerical
method automatically produces a set of discrete levels even when the
s.p. energies are larger than zero, i.e. in the continuum region.  We
did not find general criteria for the stability of our results. This
problem is not very important in the phenomenological RPA approach,
since the effects related to the truncation of the s.p. configuration
space are taken into account by changing the parameters of the
force. However, in self-consistent calculations this is a more
serious problem, since the interaction parameters are chosen to
reproduce, in HF calculations, the ground state properties of the
various nuclei, and the force remains unchanged in RPA calculations.

To keep under control these problems, we restricted our study to the
low-lying excited states. For each excitation studied we used the same
criterion considered in the phenomenological approach, i.e. we
controlled that the energy eigenvalues of the first low-lying states
did not vary by more than 0.1 MeV against the enlargement of the
configuration space. In order to obtain this numerical accuracy, we
had to use configuration spaces composed by a few thousands
s.p. states, more than 2000 in \lead. The numerical stability of
higher energy excitations, such as giant resonances, requires even
larger configuration spaces.  In this case we believe it is necessary
to abandon the discrete RPA calculations and treat correctly the
continuum, as it is done, for example, in Refs. \cite{don08t,don11}.

We have built two new forces by adding to the D1S \cite{ber91} and D1M
\cite{gor09} parameterizations of the Gogny interaction \cite{dec80} a
tensor-isospin term similar to that given by Eq. (\ref{eq:vtensor}).
We label D1ST and D1MT these new interactions. Since in Gogny-like
forces the spin-orbit term is fixed to reproduce the experimental
splitting of the s.p. energies of the $1p_{1/2}$ and $1p_{3/2}$
neutron states in \oxy, we used this nucleus as reference. The
new interactions have been fixed by using an iterative procedure. We
started with a HF calculation without tensor force to produce a set of
s.p. energies and wavefunctions to be used in the RPA
calculations. Then, we made a RPA calculation with the tensor force
and we fixed the value of the parameter $b$ of Eq. (\ref{eq:vtensor})
in order to reproduce the energy of the first $0^-$ excitation of \oxy
\/ at 10.6 MeV.  With this new interaction we recalculated the HF
s.p. energies by changing the spin-orbit interaction to reproduce the
splitting of the two $p$ states quoted above. These HF and RPA
calculations have been repeated until the convergence of the result
was obtained. By using this procedure we have found for the parameter
$b$ the value 0.6 ~fm$^{-2}$, for the D1ST force, and of 0.25
~fm$^{-2}$, for the D1MT one. Summarizing, we added a tensor term to
the D1S and D1M Gogny-like interactions and we modified only the
spin-orbit terms from 130 MeV, in the original D1S force, to 134 MeV,
in the D1ST interaction, and from 115 MeV, in the D1M force, to 122.5
MeV, in the D1MT interaction. No other values of the force parameters
have been changed.

The tensor terms of the D1ST and D1MT interactions are indicated in
panel (b) of Fig. \ref{fig:tensq} by the dashed-dotted and
dashed-doubly-dotted lines respectively. In this figure they are
compared with the tensor-isospin term of the microscopic Argonne V18
interaction (solid curve). By construction, the effective tensor terms
are smaller than that of the bare N-N interaction.  More interesting
is the comparison with the long dashed line which has been produced by
multiplying the bare interaction with the scalar part of the
short-range correlation function obtained in Correlated Basis Function
calculations \cite{ari07}. The remarkable difference between this line
and those of the D1ST and D1MT forces indicates that our procedure
includes in the effective tensor term not only the effects of the
short-range correlations, but also some other many-body effects that
the microscopic calculations consider explicitly.  In the same panel
we make a comparison with other two tensor terms of finite-range
interactions used in the literature, the GT2 \cite{ots06} and the
M3YP2 \cite{nak03} forces. The tensor term of the GT2 is constructed
to have the same volume integral of the Argonne V18 tensor force.  The
strength of this tensor force is much larger than those of the tensor
forces we have built. On the opposite, we observe that the tensor term
of the M3YP2 force is much smaller.

Even though the D1ST and D1MT forces reproduce the experimental value
of the excitation energy of the $0^-$ state at 10.6 MeV in $^{16}$O,
they produce rather different RPA wave functions, as we have verified
by calculating transition densities and inclusive neutrinos cross
sections for this excited state.

\section{Hartree-Fock results} 
\label{sec:gs}

We used the D1ST and D1MT interactions, whose construction has been
described in the previous section, to make spherical HF calculations
for a set of nuclei in different regions of the isotope chart. We have
chosen nuclei where the s.p. states below the Fermi level are fully
occupied to avoid deformation problems and to minimize pairing
effects.  We have verified these features by controlling the results
given by the deformed Hartree-Fock-Bogolioubov calculations of
Ref. \cite{del10}.  

\begin{figure}
\begin{center}
\includegraphics[scale=0.4]{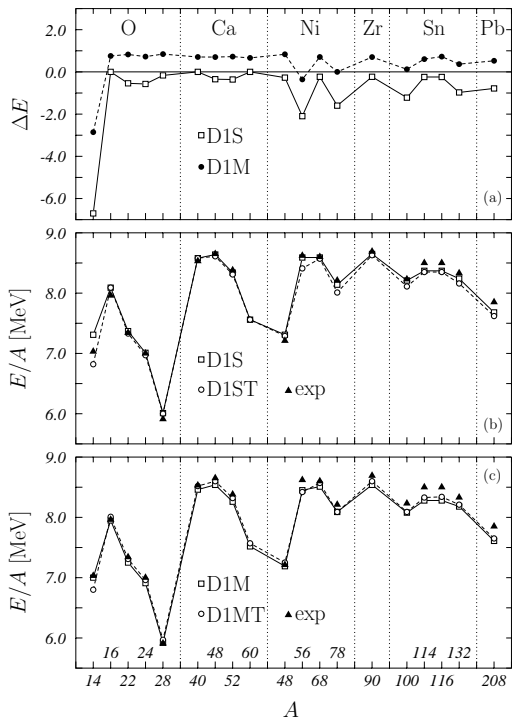} 
\caption{\small Panel(a): relative percentage differences,
  Eq. (\ref{eq:deltae}), between binding energies calculated with the
  D1ST and D1S forces (open squares) and with the D1MT and D1M (solid
  circles) interactions. Panels (b) and (c): binding energies per
  nucleon calculated with the various interactions compared with the
  experimental values (solid triangles) taken from
  Ref. \cite{aud03,audw}.  The experimental values for the
  $^{28}$O, $^{60}$Ca, $^{48}$Ni, $^{78}$Ni and $^{100}$Sn nuclei are
  estimated \cite{audw}.  The lines have been drawn to guide the eyes.
  }
\label{fig:be}
\end{center}
\end{figure}

In Fig. \ref{fig:be} we compare the binding energies obtained
for various nuclei we have considered, with the experimental values
taken from Refs. \cite{aud03,audw}. The experimental energies of the
$^{28}$O, $^{48}$Ni, $^{60}$Ca, $^{78}$Ni and $^{100}$Sn nuclei have
not been measured but estimated \cite{audw}.

The methodology used in our study is already evident.  We have
investigated the effects of the tensor force by comparing the results
obtained by using forces with and without tensor term. To emphasize
the effects of the tensor force, we show in the panel (a) of
Fig. \ref{fig:be} the quantity 
\beq
\Delta E \,= \, 100 
\, \frac{E_{{\rm D1}\alpha{\rm T}} \,
 - \, E_{{\rm D1}\alpha}}{E_{{\rm D1}\alpha}}
\, ,
\label{eq:deltae}
\eeq 
which is the relative percentage differences between the binding
energies calculated by using interactions with and without tensor
term, $E_{{\rm D1}\alpha{\rm T}}$ and $E_{{\rm D1}\alpha}$
respectively ($\alpha \equiv {\rm S,M}$). In this panel the open
squares indicate the results obtained with the D1ST and D1S
interactions and the solid circles those obtained with the D1MT and
D1M ones. The lines have been drawn to guide the eyes.

If we exclude the anomalous values obtained for the $^{14}$O nucleus,
we observe that all the other results lie in the range $\Delta E
\approx \pm 2$. In general, the inclusion of the tensor term in the D1M
force produces more binding, while it has opposite effect in the D1S
case. These results show that effect of the tensor on the binding
energy of the nuclei we have investigated is rather small, confirming
the results of Ref. \cite{co98b}. As a consequence, and as we can see
in panels (b) and (c), the inclusion of the tensor term does not
modify the agreement with experimental data in a significant manner.

\begin{figure}
\begin{center}
\includegraphics[scale=0.4]{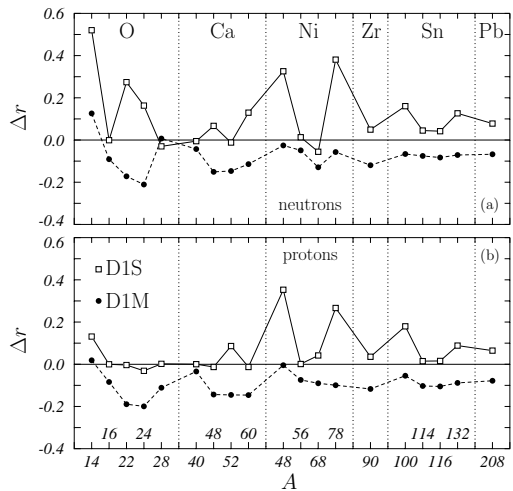} 
\caption{\small Relative percentage 
difference, Eq. (\ref{eq:deltar}), between
root mean square radii of neutrons (upper panel) and protons (lower
panel) distributions calculated with and without tensor for D1ST and
D1S (open squares) and D1MT and D1M (solid circles) interactions. 
}
\label{fig:radii}
\end{center}
\end{figure}

We have investigated the effect of the tensor force on the proton and
neutron density distributions obtained from our HF calculations, and,
also in this case, we found small effects. We summarize the results of
this study in Fig. \ref{fig:radii} where we have shown the relative
percentage difference between root mean square (rms) radii obtained by
using interactions with and without tensor force,
\beq
\Delta r \,= \, 100 \, \displaystyle \frac
{\sqrt{\langle  r^2 \rangle_{{\rm D1}\alpha{\rm T}}}\, 
  - \,\sqrt{\langle  r^2 \rangle_{{\rm D1}\alpha}} }
{\sqrt{\langle  r^2 \rangle_{{\rm D1}\alpha}}} \, . 
\label{eq:deltar}
\eeq 

We show the results for the neutron radii in the upper panel of the
figure, and those for the proton radii in the lower panel.  We have
indicated with the open squares the results obtained with the D1S
forces and with the solid circles those obtained with the D1M
ones. 

The effects of the tensor force are rather small, of the order of few
parts on a thousand. We observe again the different sign between D1S
and D1M results. The larger binding produced by the tensor in the D1M
case generates more compact nuclei, i.e. with smaller rms radii. In
the case of the D1S forces the effect is just the opposite. This trend
is present in both neutron and proton cases.

The results we have presented so far indicate that the bulk properties
of the nuclear ground states are not greatly affected by the presence
of the tensor force. The situation changes when the s.p. energies are
considered.  In the remaining part of this section we shall discuss
results concerning s.p. properties. Henceforth we shall distinguish
the proton and neutron s.p. levels by using the $\pi$ and $\nu$
labels, respectively.

A first quantity we have studied is the difference between the
s.p. energies of spin-orbit partner levels
\beq
s \, = \, \epsilon_{l-1/2} - \epsilon_{l+1/2} \, .
\eeq
In particular, we have studied the difference between the values of
$s$ obtained by using forces with and without tensor terms
\beq
\Delta s \,
= \, s_{{\rm D1}\alpha{\rm T}} \, - \, s_{{\rm D1}\alpha}
\, .
\label{eq:deltaep}
\eeq 

\begin{figure}
\begin{center}
\includegraphics[scale=0.4]{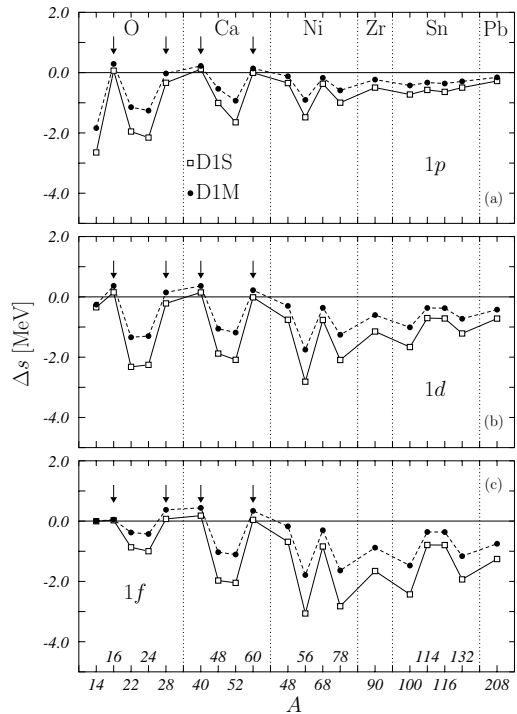} 
\caption{\small Differences between the s.p. energy differences of
  spin-orbit partners levels, Eq. (\ref{eq:deltaep}), calculated with
  and without tensor forces, for the $1p$, panel (a), $1d$,  
  panel (b), and $1f$, panel (c), proton states. The results for D1S and
  D1ST interactions are shown by open squares and solid circles,
  respectively. The arrows indicate those isotopes where the effect of
  the tensor is expected to be zero.  }
\label{fig:lsprot}
\end{center}
\end{figure}

\begin{figure}
\begin{center}
\includegraphics[scale=0.4]{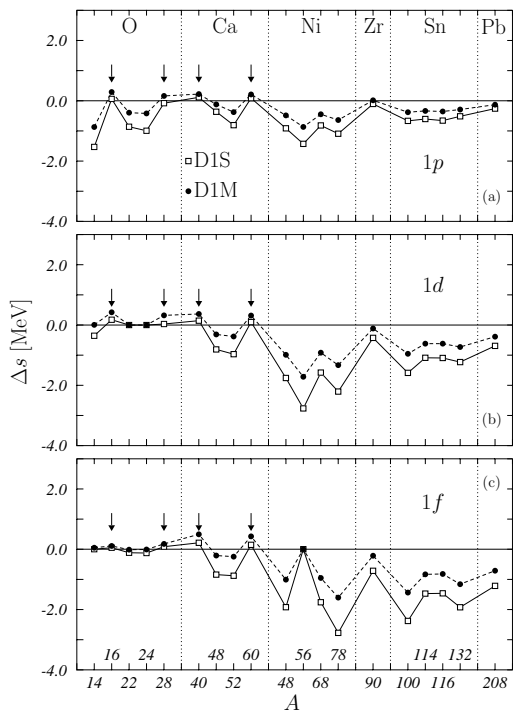} 
\caption{\small The same as Fig. \ref{fig:lsprot} for neutron states.
} 
\label{fig:lsneut}
\end{center}
\end{figure}

In Figs. \ref{fig:lsprot} and \ref{fig:lsneut} we show the values of
$\Delta s$ calculated for the $1p$, $1d$ and $1f$ proton and neutron
levels, respectively, for all the nuclei considered. In these figures
the open squares indicate the results obtained with the D1S
interactions, and the solid circles those obtained with the D1M ones.
The arrows indicate those nuclei where all the spin-orbit partner
levels, for both protons and neutrons, are fully occupied.

Let us consider first Fig. \ref{fig:lsprot}, where, for each isotope chain
we show the evolution of $\Delta s$ values with the increasing number
of neutrons.  We first observe that, for each s.p. level
investigated, the D1S and D1M results have identical behaviour. Minima
and maxima are in the same position for both type of calculations.  A
second observation is that, in general, $\Delta s$ is negative.  This
means that the tensor force reduces the energy difference between
spin-orbit partner levels. A third observation is that the effects of
the tensor force are minimal, almost zero, for those nuclei indicated
with the arrows.

The first observation indicates that the effects we have pointed out
are strictly related to peculiarities of the tensor force and of the
nucleus investigated. The small difference between the D1S and D1M
results reflects the difference between the tensor forces in D1ST and
D1MT, as we have shown in Fig. \ref{fig:tensq}.

The second and third observations are well understood within the
scheme proposed by Otsuka {\it et al.}\/ \cite{ots05,ots06,ots08}. The
effect of the tensor interaction between a proton and a neutron
occupied s.p. levels is attractive if one of the levels has an angular
momentum $j_> \equiv l+1/2$ and the other one $j_< \equiv l-1/2$. If
the angular momenta of the two s.p. levels are of the same type,
i.e. both $j_>$ or both $j_<$, the effect of the tensor force has
opposite sign.  The effect of an occupied neutron level with angular
momentum $j_>$ is to increase the energies of the s.p. proton levels
with $j_>$, and to lower those of the levels with $j_<$. As a
consequence, the splitting between the energies of the proton
spin-orbit partner levels is reduced. This effect is reversed when the
occupied neutron level has $j_<$. If both $j_>$ and $j_<$ neutron
levels are occupied, the two effects cancel with each other.

The results presented in Fig. \ref{fig:lsprot} are well explained
within this picture.  In the nuclei marked with the arrows, all the $j_>$
and $j_<$ neutron levels are occupied. In these nuclei we do not
expect any tensor effect. In reality the values of $\Delta s$ are not
exactly zero even in these cases because we have changed the strengths
of the spin-orbit interactions in the forces with tensor terms (see
Sec. \ref{sec:int}). In all the other nuclei there is, at least, one
occupied neutron level with $j_>$, whose spin-orbit partner level is
empty. The effect we have discussed above predicts negative values of
$\Delta s$, as those shown in the figure.

In Fig. \ref{fig:lsneut} we have shown the values of $\Delta s$ for
the neutron  $1p$, $1d$ and $1f$ levels. Since in the oxygen and
calcium isotopes all the proton s.p. partner  levels below the Fermi
surface are completely occupied, the effects we observe in these
isotopes are due to the tensor interaction acting within neutron
s.p. levels only. This is exactly the effect we have discussed above,
but acting between s.p. states with the same isospin. This effect is
weaker than that between states with opposite isospin, as the results
for the $^{22}$O, $^{24}$O, $^{48}$Ca and $^{52}$Ca nuclei
indicate. In these nuclei one of the neutron spin-orbit partner levels
is unoccupied and the tensor effect is expected to be
present. Actually, there is also an effect produced by the different
values of the spin-orbit forces, but we have verified that, for the
cases under consideration, this effect is negligible.

The analysis of the $\Delta s$ results for the Ni, Zr, Sn and Pb
isotopes is more complicated, since, in these cases, in both the
proton and neutron sectors, there are spin-orbit partner s.p. levels
not fully occupied. For this reason, we must consider the interaction
with both proton and neutron levels.  The similarity between the
behaviour of the results obtained with the different interactions
indicate that we are observing an effect due to the peculiarities of
the tensor force.

\begin{figure}
\begin{center}
\includegraphics[scale=0.4,angle=90]{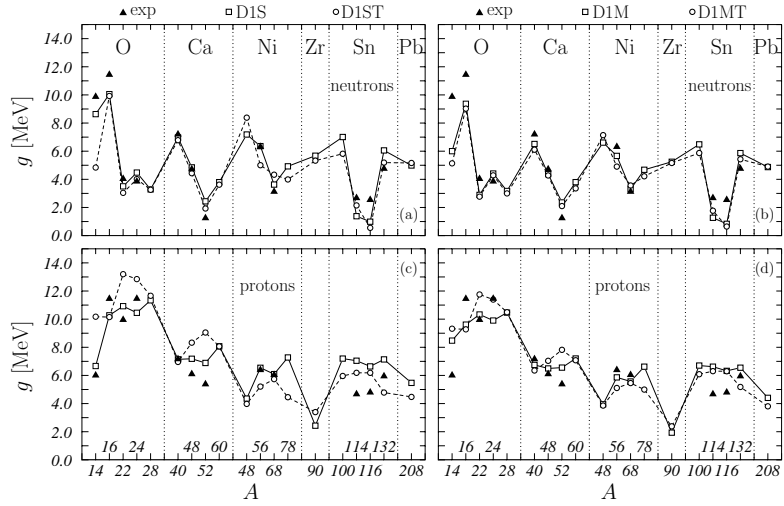} 
\caption{\small Neutron, panels (a) and (b), and proton,  
  panels (c) and  (d), energy gaps, 
  in MeV, for the various nuclei and interactions we have
  investigated. In the left panels we show the results obtained by
  using the D1S interactions and in the right panels those obtained
  with the D1M forces.  The experimental values, solid triangles, have
  been extracted from the binding energies of the neighbouring
  nuclei \cite{aud03,audw}.  }
\label{fig:gap}
\end{center}
\end{figure}

Another quantity of interest related to the s.p. energies is the gap,
$g$, between the energies of the levels just above and just below the
Fermi surface. In Fig. \ref{fig:gap} we have shown the proton (lower
panels) and neutron (upper panels) energy gaps calculated for the
nuclei we have investigated with the four forces we are using.  Our
results are compared with the experimental values (solid triangles)
extracted from the binding energies of nuclei with atomic numbers
differing by one unit.

The first remark related to the results shown in Fig. \ref{fig:gap} is
that, in general, the effects of the tensor force are relatively small
and they are similar for both type of interactions.  They are
negligible for neutrons, while some noticeable effects are present in
the case of protons. The neutron results, obtained with and without
tensor, follow reasonably well the behaviour of the experimental
energy gaps.

\begin{figure}[b]
\begin{center}
\includegraphics[scale=0.4]{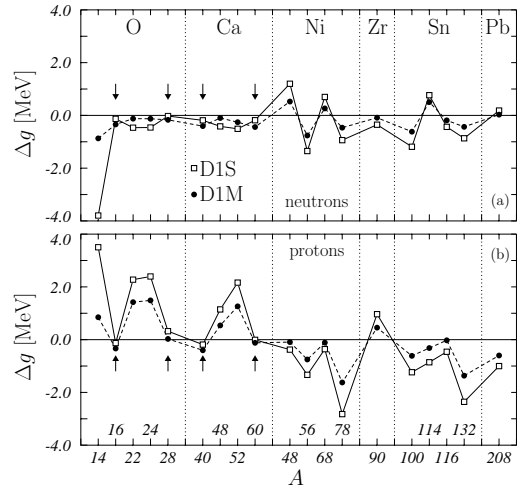} 
\caption{\small Differences between the energy gaps calculated with and
  without tensor force, 
  for neutrons, panel (a), and protons, panel (b). The results
  obtained with the D1S type interactions are indicated by the open squares, 
  and those obtained with the D1M type  interactions by the solid
  circles. The arrows indicate those isotopes where the tensor effect
  is expected to be zero.  }
\label{fig:deltagap}
\end{center}
\end{figure}
  
The effects of the tensor force are better emphasized in
Fig. \ref{fig:deltagap} where we show the difference
\beq 
\Delta g \, = \, g_{{\rm D1}\alpha{\rm T}} \, - \, g_{{\rm D1}\alpha} \, ,
\label{eq:deltag}
\eeq 
between the gap values obtained by using interactions with and without
the tensor force. Also in this case we have indicated with the arrows
the nuclei where all the spin-orbit partner levels are fully occupied,
for both protons and neutrons. The results obtained with the D1S
interactions are shown by the open squares and those obtained with the
D1M interactions by the solid circles. It is interesting to observe
that also in this case the behaviour of the D1S and D1M results is
similar, though the effects produced by the D1ST force are slightly
larger than those obtained for D1MT.

Also the results in Fig. \ref{fig:deltagap} can be well explained
within the scheme proposed by Otsuka. In the spin unsaturated oxygen
isotopes, the unpaired neutron levels are always of $j_>$ type. The
tensor force lowers the energy of the $(1p_{1/2})_\pi$ occupied proton
level ($j_<$ type) and increases that of the $(1d_{5/2})_\pi$ empty
level ($j_>$ type).  For this reason, the results for the oxygen
isotopes, in the proton case, have positive values. An analogous
effect is present also for the $^{48}$Ca and $^{52}$Ca isotopes. In
this case, the states to be considered are the holes $(1d_{3/2})_\pi$
or $(2s_{1/2})_\pi$ and the particle $(1f_{7/2})_\pi$.

For the heavier isotopes, the situation is more complicated because
the unpaired levels can be more than one, and because the levels
involved in the gap calculation could be both of the same type ($j_>$
or $j_<$).  This is the case for $Z$ or $N$=28 or 50, but not for
$Z=40$. For this reason, in the case of protons, the values of $\Delta
g$ are negative for Ni, Sn and Pb isotopes and positive for $^{90}$Zr.

For the neutrons, panel (a), the situation is more difficult to
discuss since the nuclei we have investigated are not isotones. For
each set of isotopes there are different s.p. levels related to the
neutron gap, therefore the effect of the tensor term can be different
for each nucleus considered.  This is the reason of the oscillating
behaviour we observe in the Ni isotopes.

Since oxygen and calcium isotopes are spin saturated in protons, the
results presented in the panel (a) of Fig. \ref{fig:deltagap} for
these nuclei, are produced by the interaction of an unpaired neutron
s.p. level of $j_>$ type with the neutron s.p. levels just below and
above the Fermi surface. The comparison of the results of these nuclei
shown in the two panels indicates that the effect of the tensor
interaction between like nucleons is smaller than that between
neutrons and protons, reflecting the fact that we have used a
tensor-isospin term in the interaction.

\begin{figure}
\begin{center}
\includegraphics[scale=0.4]{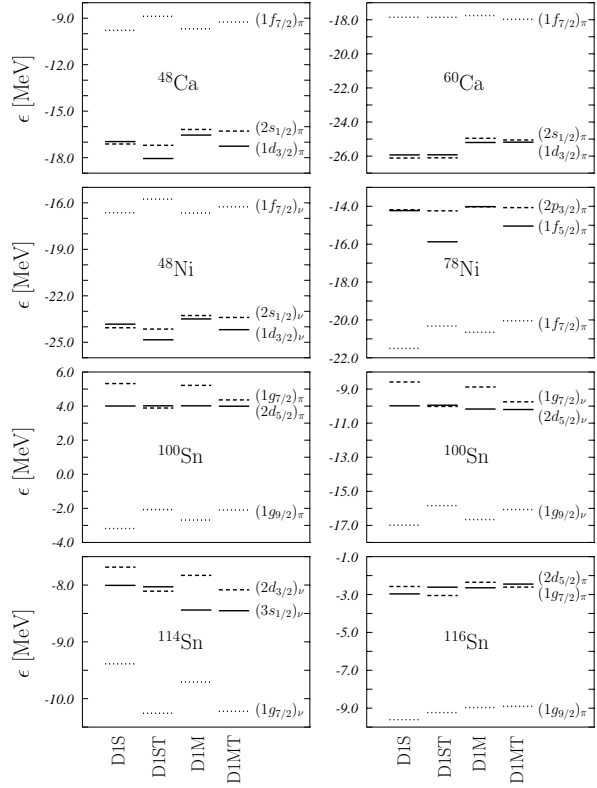} 
\caption{\small Single particle levels around the Fermi surface which
  change their order when the tensor force is used.}
\label{fig:spetens}
\end{center}
\end{figure}

We have seen that the tensor term affects more the s.p. than the bulk
properties of the nucleus. The effects of the tensor force on the
s.p. levels can modify their relative order. If this occurs for the
levels near the Fermi surface, the spin of even-odd nuclei neighboring
the nuclei investigated, which is determined by the last unpaired
nucleon, should be modified. We found some cases where this happens
and few of them are presented in Fig. \ref{fig:spetens}, where we show
the evolution of the states near the Fermi surface for the different
interactions we have used.

In the $^{48}$Ca case we consider the two proton states below the
Fermi surface. The $(1d_{3/2})_\pi$ is a $j_<$ state, therefore its
energy is lowered by the tensor, while that of the $(2s_{1/2})_\pi$
state remains essentially unchanged. This effect inverts the order of
the two states, as we observe in the results of the D1ST column. The
calculations done with the D1M interaction give the two proton 
states in an inverted order with respect to that of the D1S one. In
this case the tensor effect enlarges the energy difference between the
two states. Since $^{48}$Ca and $^{48}$Ni are mirror nuclei, we
expected an analogous effect for the two neutron hole states in
$^{48}$Ni. This effect is present in our calculations as it is shown
in the figure.  

Since all the spin-orbit partner levels are occupied for both protons
and neutrons, we do not expect tensor effects in $^{60}$Ca.  This is
what we observe in Fig. \ref{fig:spetens}. We show the results for
this nucleus since the order of the proton hole states obtained with
the D1S and D1ST interactions is inverted with respect to that
obtained with the D1M and D1MT forces. 

In the $^{78}$Ni case, the tensor force produces a large lowering of
the $(1f_{5/2})_\pi$ level, and this, for the D1M interaction,
generates an inversion with the $(2p_{3/2})_\pi$ level. The figure
shows that the tensor force reduces the energy gap between
$(1f_{5/2})_\pi$ and $(1f_{7/2})_\pi$ levels.

For the $^{100}$Sn nucleus we show both proton and neutron cases,
since in both situations we obtain the inversion of the $1g_{7/2}$ and
$2d_{5/2}$ levels, when the D1ST interaction is used. The tensor force
lowers the energies of the $1g_{7/2}$ levels, which are $j_<$ states,
and enhances those of the $2d_{5/2}$ levels, which are $j_>$ states.
We obtain a large effect on the energies of the $1g_{7/2}$ levels
while the modification of the $2d_{5/2}$ energy is minimal.

We observe an inversion of the order of the $(2d_{3/2})_\nu$ and
$(3s_{1/2})_\nu$ levels in $^{114}$Sn, when the D1ST interaction is
used. The tensor lowers the energy of the $(2d_{3/2})_\nu$ state, of
$j_<$ type, and has no effect on the $(3s_{1/2})_\nu$ state. The effect
is present also when the D1MT interaction is used, but it is not large
enough to invert the order of the states.

In $^{116}$Sn we obtain an inversion of the order of the
$(1g_{7/2})_\pi$ and $(2d_{3/2})_\pi$ levels when both D1ST and D1MT
interactions are used.  Also in this case we observe the tensor effect
predicted by the Otsuka's scheme. The energy of the $(1g_{7/2})_\pi$
state, of $j_<$ type, is lowered, while that of the $(2d_{3/2})_\pi$
state, of $j_>$ type, is enhanced.

To summarize the results presented in this section, we may say that
the tensor effects have remarkable consequences on those observables
which we relate to s.p. properties of the nuclear system, such as
s.p. energies, gaps and spin of the system.

%
\section{RPA calculations} 
\label{sec:rpa}

In the previous section we have presented the results regarding the
ground state properties of some spherical nuclei. In this section we
show the results we have obtained for the excited states of these
nuclei by doing RPA calculations. In this study, we have also
considered $^{12}$C, a well studied nucleus from both experimental and
theoretical points of view, despite the fact that its ground state
contains deformations \cite{del10}. This feature is less important
in the description of the excitation spectrum than for the ground state
observables. 

The input required by any RPA calculations is composed by the
s.p. basis and the effective interaction. The results shown in this
section have been obtained by using the s.p. basis and the effective
interaction provided by HF calculations. In the RPA jargon this
procedure is called self-consistent to distinguish it from the
phenomenological one described in Sec. \ref{sec:intro}. To be precise,
we do not strictly use the same interaction in HF and RPA, since in
the latter case we neglect the Coulomb and spin-orbit terms of the
interaction. We should keep in mind this fact, even though there are
indications that, in RPA calculations, these two terms of the
interaction produce very small effects that, in addition, have the
tendency of canceling with each other \cite{sil06}.  

The strategy of our investigation is analogous to that used in the
previous section, i.e. we compare results obtained by using
interactions with and without tensor terms. While for the HF
calculations this comparison provides a clear indication of the effects
of the tensor force, in the RPA case the situation is more
complicated. In effect, the differences in the RPA results can be due
to genuine effects of the tensor interaction in the RPA calculations,
and also to the different values of the s.p. energies and wave
functions provided by the HF results.

We have disentangled the effects produced by these two different
sources by presenting results where the tensor force is switched on
and off in both HF and RPA calculations. To distinguish the results of
the different type of calculations we have indicated with $\omega^{\rm
RPA}_{\rm ab}$ the RPA excitation energies, where the first subindex,
a, refers to the interaction used in the HF calculation, and the
second subindex, b, to that used in the RPA one. The a and b labels
can be t if the interaction includes the tensor term, i.e. if the D1ST
or the D1MT interaction has been used, and n otherwise. For example,
$\ornt$ indicates the excitation energy obtained with a HF calculation
without tensor and a RPA calculation with the tensor force. We present
also results obtained by switching off the residual interaction. We
label these results as $\omega^{\rm IPM}_{\rm a}$, where the
superscript IPM means Independent Particle Model. The energies
produced in this type of calculations need only one subindex which
indicates the presence, t, or the absence, n, of the tensor force in
the HF calculation.

In our study we have investigated various types of multipole
excitations, and, in agreement with the results of
Refs. \cite{spe80,co90,cao09}, we found that the natural parity
excitations are practically insensitive to the effects of the tensor
force. For this reason, we present here only results obtained for the
unnatural parity excitations.

We start or discussion by presenting the results related to the
excitation of the first $0^-$ state in the various isotopes we are
studying. In the phenomenological calculations of Sec. \ref{sec:int}
we have pointed out the large sensitivity of the excitation energy of
these states to the tensor force. For this reason we have chosen the
energy of this excitation in \oxy \/ to select the strengths of our
tensor forces.

%
\begin{table}[tb]
\begin{center}
\begin{tabular}{rccccc}
\hline \hline
          & exp   & D1S   & D1ST  & D1M   & D1MT \\ \hline
$^{12}$C  & 18.40 & 19.63 & 14.42 & 18.83 & 15.27 \\
$^{16}$O  & 10.96 & 13.95 & 10.94 & 13.08 & 10.96 \\
$^{40}$Ca & 10.78 & 12.22 &  9.57 & 11.56 &  9.60 \\
$^{48}$Ca &  8.05 & 14.10 & 11.63 & 12.85 & 11.26 \\
$^{208}$Pb & 5.28 &  8.27 &  7.93 &  8.24 &  7.92 \\
\hline \hline
\end{tabular}
\caption{\small Energies, in MeV, of the first $0^-$ excited state for
  those isotopes we have studied in this work where the 
  energy values have been measured \cite{heu07,bnlw}.  
  The theoretical
  energies are obtained by doing RPA calculations with different
  interactions.
  }
\label{tab:zerom}
\end{center} 
\end{table}

\begin{figure}[b]
\begin{center}
\includegraphics[scale=0.4, angle=0] {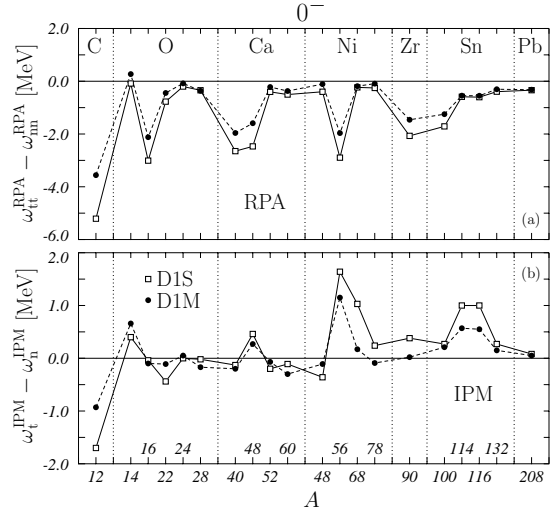} 
\caption{\small Differences between the energies of the first $0^-$
  excited state obtained with and without tensor forces, for the
  nuclei under investigation. D1S (D1M) results are plotted with open
  squares (solid circles). In the panels (a) and (b) we show the
  results obtained with RPA and IPM calculations, respectively.}
\label{fig:zerom}
\end{center}
\end{figure}

For the nuclei where the $0^-$ excitation energies have been
experimentally clearly identified \cite{heu07,bnlw}, we show in Table
\ref{tab:zerom} the excitation energies obtained in our RPA
calculations.  The tensor interaction lowers the excitation energies
in all the cases.  The perfect agreement of the \oxy \/ case is obtained
by construction, but we have improvements also in the \caI, \caII \/ and
\lead \/ cases.  The only worsening produced by the inclusion of the
tensor term is that of the \car \/ nucleus, which we know to be a
difficult nucleus to describe within our theoretical framework
tailored to spherical systems.

A general view of the tensor effects in all the nuclei we have
considered is given in Fig. \ref{fig:zerom}. In panel (a) we show the
differences $\ortt - \ornn$ for the first $0^-$ states in all the
nuclei under investigation.  The open squares show the results for the
interactions of D1S type, while the solid circles those of the D1M
one.  To separate the tensor effects in RPA calculations from those
produced by the change in the s.p. configuration space, we show in
panel (b) the energy differences $\oit-\oin$.

In the RPA calculations, the results obtained with the tensor force
are always lower than those obtained without it, and this produces
negative values of the differences $\ortt - \ornn$.  The only
exception to this general trend is that of the $^{14}$O nucleus in the
case of the D1M interaction. In reality we observe in panel (b) that
the differences $\oit - \oin$ for this nucleus are positive and larger
than those found in the RPA calculation, indicating that also in this
case the tensor force in the RPA calculation lowers the energy value
of the first $0^-$ state.

The results presented in the two panels do not show any correlation.
This indicates that the effect shown in the panel (a) of the figure is
a genuine effect of the tensor force on the RPA calculations, which, in
the case of the $0^-$ excitation, is always attractive, even in the
$^{14}$O case, confirming the results we have obtained with the
phenomenological calculations and shown in Fig. \ref{fig:ene0-}.

We have investigated the effects of the tensor force on multipole
excitations with strong isoscalar (IS) or isovector (IV) character. To
identify well these states we have considered only isotopes with
$N=Z$.  In these nuclei, we have selected those multipole excitations
composed mainly by two identical particle-hole pairs in terms of
angular momentum coupling, but one for protons and the other one for
neutrons. This is the ideal situation to produce IS and IV partners
levels. In the first case proton and neutron excitations are in phase,
while in the second one are off phase. In our RPA results the IS and
IV partner states are easily identifiable by observing the relative
phases of the RPA amplitudes of the main particle-hole pairs.

\begin{table}[htb]
\begin{center}
\begin{tabular}{rccccccccccccccccc}
\hline\hline 
      &~~~~~ & &~~~~~&\multicolumn{2}{c}{D1S} &~~~~~
                 & \multicolumn{2}{c}{D1ST} &~~~~~
                 & \multicolumn{2}{c}{D1M} &~~~~~
                 & \multicolumn{2}{c} {D1MT} &~~~~~
                 & \multicolumn{2}{c} {exp}\\\cline{5-6}\cline{8-9}\cline{11-12}\cline{14-15}\cline{17-18}
      &    & $J^\pi$ &&  IS & IV && IS & IV && IS & IV && IS & IV && IS & IV \\ \hline
    $^{ 12}$C && $1^+$ &&    4.78 &    7.71 &&    1.94 &    8.17 &&    3.44 &    7.21 &&    2.43 &    7.68 && 12.71 &  15.11\\
    $^{ 12}$C && $2^-$ &&   15.75 &   18.62 &&   14.27 &   17.30 &&   14.59 &   18.06 &&   13.50 &   17.15 && 11.83 &  13.35\\
    $^{ 12}$C && $4^-$ &&   17.11 &   18.35 &&   15.61 &   17.08 &&   16.22 &   17.40 &&   15.18 &   16.61 && 18.27 &  19.50\\
    $^{ 16}$O && $2^-$ &&   10.36 &   12.10 &&    9.64 &   12.08 &&    9.47 &   11.26 &&    8.74 &   11.08 &&  8.87 &  12.53\\
    $^{ 16}$O && $4^-$ &&   17.08 &   18.20 &&   16.52 &   17.96 &&   16.02 &   16.97 &&   15.63 &   16.86 && 17.79 &  18.98\\
   $^{ 40}$Ca && $2^-$ &&    7.61 &    8.98 &&    6.62 &    8.83 &&    6.63 &    8.31 &&    5.82 &    8.01 &&  7.53 &   8.42\\
   $^{ 40}$Ca && $4^-$ &&    6.93 &    7.51 &&    6.37 &    7.41 &&    6.68 &    7.04 &&    6.02 &    6.70 &&  5.61 &   7.66\\
   $^{ 40}$Ca && $6^-$ &&   14.48 &   15.15 &&   14.15 &   14.95 &&   13.66 &   14.17 &&   13.44 &   14.07 && &\\
   $^{ 56}$Ni && $2^-$ &&   11.64 &   14.32 &&   11.06 &   13.91 &&   11.10 &   13.50 &&   10.62 &   13.35 && &\\
   $^{ 56}$Ni && $4^-$ &&   12.57 &   13.60 &&   11.98 &   13.08 &&   12.01 &   12.93 &&   11.58 &   12.56 && &\\
   $^{ 56}$Ni && $5^+$ &&    6.73 &    7.13 &&    4.86 &    5.37 &&    6.00 &    6.29 &&    5.00 &    5.52 && &\\
   $^{100}$Sn && $3^+$ &&    8.70 &    8.97 &&    5.56 &    6.38 &&    8.09 &    8.21 &&    6.37 &    6.68 && &\\
   $^{100}$Sn && $5^+$ &&    7.10 &    7.39 &&    5.53 &    6.29 &&    6.66 &    6.88 &&    5.97 &    6.22 && &\\
   $^{100}$Sn && $7^+$ &&    7.30 &    7.56 &&    5.44 &    6.18 &&    7.91 &    8.05 &&    6.03 &    6.38 && &\\
\hline \hline 
\end{tabular}
\caption{\small Excitation energies, in MeV, in nuclei with $Z=N$,
    for different multipoles,
    where we have identified isoscalar (IS) and isovector (IV)
    character. The experimental values are taken from Ref.\cite{led78}.}
\label{tab:isiv}
\end{center} 
\end{table}

We have presented in Table \ref{tab:isiv} the energies of the excited states
we have investigated, and we compare them with the available
experimental values \cite{led78}.
All the results indicate that the energies of
the IV excitations are larger than those of the IS ones, as it is
experimentally well established. In Ref. \cite{don09} we obtained for
the calculations done with the D1S interaction an opposite
behaviour. These results were wrong, since we found an error in the
treatment of the exchange part of the density dependent terms of the
D1S force.

\begin{figure}[b]
\begin{center}
\includegraphics[scale=0.4,angle=90] {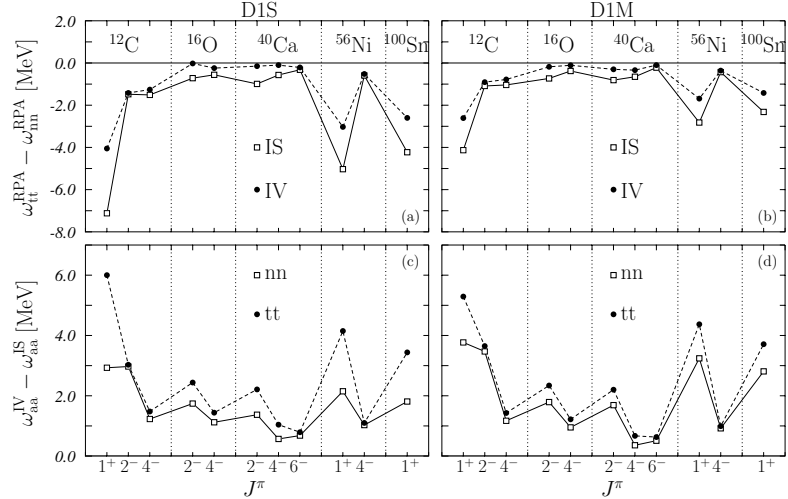} 
\caption{\small In the panels (a) and (b) we show the differences
  between the energies calculated with and without tensor terms for
  the IS and IV states listed in Table \ref{tab:isiv}. In the panels
  (c) and (d) we show the differences between the energies of the IV
  and IS states obtained in fully self-consistent RPA calculations 
  with and without tensor forces. 
  In the panels (a) and (c) we show the results obtained with the D1S
  type forces, and in the other two panels those obtained with 
  the D1M type forces.}
\label{fig:isiv}
\end{center}
\end{figure}
\begin{figure}[h]
\begin{center}
\includegraphics[scale=0.4] {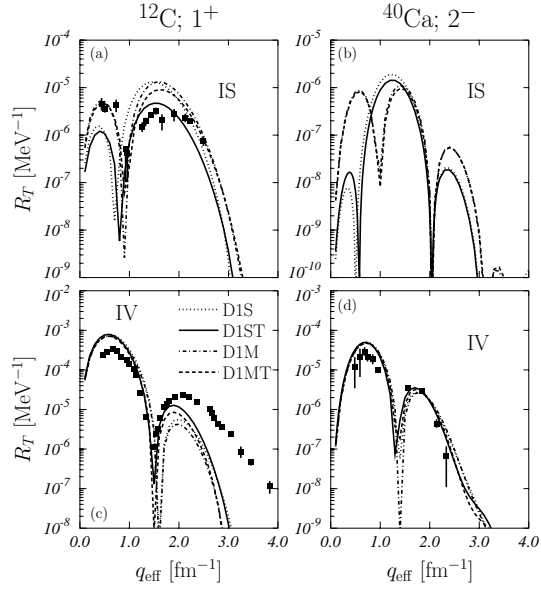} 
\caption{\small Inelastic electron scattering transverse 
  response as a function
  of the effective momentum transfer \cite{hei83}, calculated by using
  RPA wavefunctions obtained in fully self-consistent calculations done 
  with different interactions. The data are from
  Refs. \cite{hyd84t,wil87}.  
  }
\label{fig:caca}
\end{center}
\end{figure}

The effects of the tensor terms are better presented in
Fig. \ref{fig:isiv}. In the panels (a) and (b) we show the energy
differences $\ortt-\ornn$ for IS (open squares) and IV (solid circles)
states. The results found for the various calculations have rather
similar behaviours. The tensor effects are smaller on the IV states.
In the panels (c) and (d) we show the differences between the energies
of the IV and IS states for each multipole we have considered. In
these panels the open squares show the results obtained without tensor
interaction, while the solid circles include the tensor in both HF and
RPA calculations. The tensor force always increases these
differences. In general, this enhancement is larger for D1S than for
D1M and it is worth pointing out that the effect is relatively large
for the three $1^+$ states studied (in $^{12}$C and $^{56}$Ni and
$^{100}$Sn) in case of the D1S interaction.

We have studied the effects of the tensor force on the electron
scattering responses. A detailed presentation of these results would
require a discussion for each specific excited state. We plan to make
this discussion in the future. At the moment we can summarize the main
and general feature we have observed by saying that the effects of the
tensor force are larger on the IS excitations than on the IV ones. As
example, we show in Fig. \ref{fig:caca} the inclusive inelastic
electron scattering transverse responses \cite{hei83} as a function of
the effective momentum transfer, for the 1$^+$ isospin excitation
doublet in \car, and the 2$^-$ doublet in \caI, and we compare them
with the available experimental data \cite{hyd84t,wil87}.  It is
evident that the effects of the tensor are greater on the IS than on
the IV excited states.

The explanation of these facts is related to the structure of the
electromagnetic excitation of the unnatural parity states, which is
dominated by the magnetization current \cite{hei83}. The magnetization
current depends on the anomalous magnetic moment of the nucleon, which
has different sign for protons and neutrons. Since in the IS
excitations the main proton and neutron RPA amplitudes have the same
sign, the proton and neutron magnetization currents subtract with each
other. In the IV excitations the effect is reversed. Small differences
in the proton and neutron structure of the RPA wavefunctions are
emphasized by the difference and hidden in the sum of the proton and
neutron contributions.

\begin{figure}[t]
\begin{center}
\includegraphics[scale=0.4] {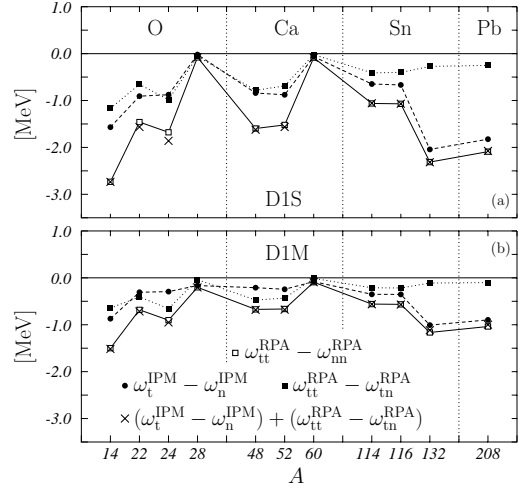} 
\caption{\small Differences between the energies of the 1$^+$ excitations
  having the largest $B(M1)$-value, in various nuclei. Solid circles
  indicate the results obtained in the IPM by using s.p. wave
  functions and energies obtained in HF calculations done with and
  without tensor. The solid squares show the results obtained in RPA
  calculations with and without tensor force, using the s.p. wave
  functions and energies obtained in HF with tensor force. The crosses
  show the sum of the two previous results. Open squares show
  the energy differences obtained by doing complete self-consistent
  calculations where the tensor force is used or not in both HF and
  RPA calculations.}
\label{fig:FFF1+}
\end{center}
\end{figure}

A remarkable sensitivity to the tensor force of the 1$^+$ excitation
of \caII \/ has been pointed out in Ref. \cite{cao09}.  The two tensor
forces used together with the Skyrme interactions in that article
produce opposite results. We have investigated how our tensor forces
affects the energy of the first 1$^+$ excitation in the nuclei with
different number of protons and neutrons we have considered. We have
summarized in Fig. \ref{fig:FFF1+} our results. In this figure we show
the differences between the 1$^+$ excitation energies obtained in
different type of calculations. The states selected are those with the
largest $B$-value and all of them are dominated by a neutron p-h
configuration. Solid circles indicate the energy differences
$\oit-\oin$. With the black squares we have shown the energy
differences $\ortt - \ortn$ indicating the effect of the tensor forces
in the RPA calculation. Finally, the open squares show the energy
differences $\ortt-\ornn$, which are linked to the global effect of
the tensor in our self-consistent calculations.

The results shown in the figure indicate that the presence of the
tensor force changes the s.p. energies in a way that the energy of the
excitation is reduced. This is shown by the fact that all the energy 
differences have negative sign. These results are in agreement with
the findings of Cao {\it et al.} \cite{cao09} for the T44 interaction
\cite{les07} and they have opposite behaviour of those found by the
same authors for the modified SLy5 interaction \cite{col07}. In their
calculations this is due to the change of the overall sign of the
tensor term for the N-N pairs with the same isospin
($\pi\pi$ or $\nu\nu$ pairs).

As we have observed in all the results presented so far, the tensor
effects are smaller for the D1M interaction.  The effect of the tensor
force is almost zero for the two nuclei where all the s.p. spin-orbit
partner levels are occupied, i.e. $^{28}$O and $^{60}$Ca.  With the
exception of these two cases, the differences $\ortt-\ornn$ are
noticeably larger than $\oit - \oin$.

To investigate the effect of the tensor force in RPA, we have
calculated the energy differences $\ortt-\ortn$ where the same
s.p. basis is used for the two RPA calculations. These results are
presented in Fig.  \ref{fig:FFF1+} by solid squares. We added
incoherently the $\oit - \oin$ and the $\ortt-\ortn$ results and we
obtained the results plotted with crosses which reproduce very well
the results of the complete calculation $\ortt-\ornn$. These results
validate the assumption used by Cao {\it et al.} in Ref. \cite{cao09}.

We conclude this section by presenting the results obtained for the
1$^+$ excitation of \lead \/ which has attracted great attention in the
past \cite{las85,mul85,tai87,las88}, and whose interest has been
renewed by recent measurements \cite{shi08}. Energies and $B$-values
for the first two $1^+$ excited states obtained with the four
interactions are shown in Table \ref{tab:pb1+}. We observe that the
tensor lowers the values of the energies of both states. The $B(M1)$
value of the first state is lowered by the tensor force, while that of
the second excited state, which is the state included in
Fig. \ref{fig:FFF1+}, is increased.

%
\begin{table}[h]
\begin{center}
\begin{tabular}{rccccc}
\hline \hline
                            & exp  & D1S  & D1ST& D1M   & D1MT \\ 
\hline
$E(1^{+}_1)$ [MeV]          & 5.85 & 7.80 & 4.76 & 6.50 & 4.82 \\
$B(M1)_1$   [$\mu_{\ n}^2$] & 2.0  & 5.08 & 2.41 & 2.33 & 1.80 \\ 
\hline
$E(1^{+}_2)$ [MeV]       & 7.1-8.7      & 10.15 & 8.06& 9.42 & 8.38 \\
$B(M1)_2$   [$\mu_{\ n}^2$]& 16.0 (17.9)& 29.63 &32.84&31.46 &32.26  \\
\hline \hline
\end{tabular}
\caption{\small Energies and $B(M1)$ values, expressed in terms of nuclear
  magnetons, of the first two
  $1^+$ excitations in \lead \/ obtained by using different interactions.  
  The experimental values are taken from  Ref. \cite{shi08}.
  }
\label{tab:pb1+}
\end{center} 
\end{table}

The detailed analysis of the results shows that the effects of the
tensor force on the energy values improve the agreement with the
experimental values. The situation regarding the $B(M1)$ values is
more complex. The tensor force lowers the $B(M1)$ values of the first
state, and this improves the agreement with the experiment. The
situation is reversed for the second excited state. Our results are
compatible with those found in literature
\cite{spe80,las85,co90,shi08}. The large difference between the
theoretical and experimental $B(M1)$ values of the second state can be
attributed to the limitations of the RPA which is unable to describe
the large fragmentation of the second 1$^+$ state.

\begin{figure}
\begin{center}
\includegraphics[scale=0.4] {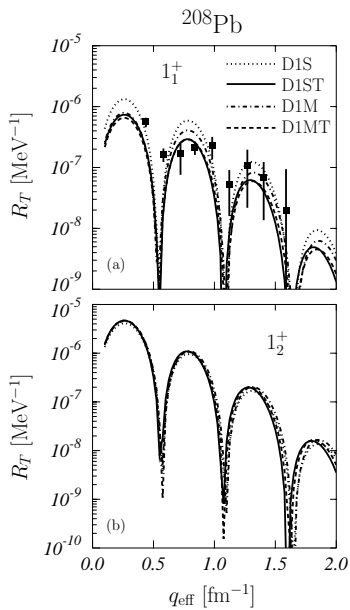} 
\caption{\small 
  Inelastic electron scattering transverse responses as a function 
  of the effective momentum transfer \cite{hei83}, calculated by using
  RPA wavefunctions obtained in fully self-consistent calculations
  done with different interactions. In panel (a)
  we show the results obtained with the RPA wavefunctions of the
  lowest excited states. The RPA wave functions used to obtain the
  results shown in panel (b) are those of the second excited state.
  The experimental data are from Ref. \cite{mul85}. 
   }
\label{fig:pb1+}
\end{center}
\end{figure}

We show in Fig. \ref{fig:pb1+} the inelastic electron scattering
responses \cite{hei83}, calculated for these two states and we 
compare them with the data of Ref. \cite{mul85}. We observe that
there is a larger sensitivity to the tensor force in the case of the
first excited state.

\section{Summary and conclusions} 
\label{sec:conc}
We have studied the effects of the tensor terms of interactions used
in HF and RPA theories.  We considered only finite-range terms in the
tensor-isospin channel, because they are the most important terms in
microscopic interactions.  We constructed these tensor terms by
multiplying the analogous term of the microscopic Argonne V18
interaction by a function which modifies its behaviour at short
internucleonic distances. This function contains a single parameter
whose value determines the strength of the tensor force.

To  determine  this strength,  we  searched  for  a global  observable
particularly   sensitive  to   the   tensor  force,   and,  by   using
phenomenological RPA  calculations, we found  it in the  excitation of
the  $0^-$  states.  We  have  also  observed  that the  tensor  force
strongly influences  the values of the s.p.  energies.  We constructed
new effective  interactions by adding a tensor-isospin  term to finite
range interactions,  and we  selected the strength  of the  tensor and
spin-orbit forces to reproduce,  in the \oxy \/ nucleus, the experimental
energy of the first $0^-$ excited state, and the splitting between the
s.p.  energies of  the neutron  $1p_{3/2}$ and  $1p_{3/2}$  levels. We
based  our work  on the  D1S and  D1M parameterizations  of  the Gogny
interaction,  and  we called  D1ST  and  D1MT  the two  new  effective
interactions constructed  with a recursive procedure where  HF and RPA
calculations  have been  repeated until  both  experimental quantities
have been reproduced.

The study of the effects of these tensor forces has been done by
comparing the results obtained in HF and RPA calculations, using
interactions with and without tensor terms. We have repeated each
calculation with both D1S and D1M type of effective interactions to
extract genuine tensor effects from those related to the peculiarities
of a specific choice of the force parameters. We always found great
similarities between the results obtained with the two different type
of parameterizations. Since the strength of the tensor force in D1MT
is weaker than that of D1ST, we found smaller tensor effects in the
results obtained with D1MT than in those for D1ST.  We have done
calculations for a set of spherical nuclei chosen such as all the
s.p. levels below the Fermi surface are fully occupied. With this
choice we avoided the effects of the deformation and we minimized
those of the pairing.

Our HF calculations indicate that tensor forces do not produce
sensitive effects on bulk observables such as binding energies, radii
and density distributions.  The effects of the tensor force on
quantities related to the s.p. properties of our theoretical approach
are certainly more remarkable. We have calculated energy splitting
between spin-orbit partner levels, energy gaps between the s.p. states
just above and below the Fermi surface, and we found noticeable
effects produced by the tensor force. Also the ordering of the
s.p. levels around the Fermi surface, which determines the spin of the
even odd nuclei neighboring those we have studied, is strongly
influenced by the presence of the tensor force in some cases. We could
explain all our results within the picture proposed by Otsuka and
collaborators \cite{ots05,ots06}, eventually by extending it to
consider the interaction between nucleons of the same type.

We have already mentioned in the introduction that the clear
identification of the tensor effects on the s.p. properties from the
comparison with experiment is rather problematic. In the MF picture
the effects of the tensor force are of the same size of those produced
by the spin-orbit force. Beyond the MF model, it is well known that
s.p. properties are strongly affected by correlation effects, the most
important ones, for the quantities of our interest, are those related
to the coupling of s.p. states with low-lying collective vibrations
\cite{bor10}.  

The study of tensor effects in the excitation spectra is more complex,
since the effects of the tensor force in the RPA calculations add to
those already present in the HF calculation that produces the
s.p. bases. In our investigation we have disentangled the effects
coming from these two different sources.  

We have verified the well know fact \cite{spe80,co90} that natural
parity excitations are essentially unaffected by the tensor force. For
this reason we have presented results regarding unnatural parity
excitations only. We started our investigation by studying the $0^-$
excitations in various nuclei, and we found that the excitation energies
obtained with the tensor forces are always smaller than those obtained
without it. 

We studied the different role played by the tensor force in IS and IV
type of excitations. To identify clearly these different type of
excitation modes, we have considered nuclei with equal number of
protons and neutrons, and states dominated by particle-hole transitions
with the same angular momentum coupling for both protons and neutrons.
The energies of the IV modes are always greater than those of the IS
modes with or without tensor. These results agree with the
experimental observations. We found that the IS excitations are more
sensitive to the tensor force than the IV ones. The tensor force
increases the energy difference between IS and IV excitations.

We made a systematic study of the effects of the tensor force on the
excitation of the 1$^+$ states in nuclei with different number of
protons and neutrons. We have considered the states showing the
largest $B(M1)$-values, and we found that the tensor force
consistently lowers the values of their excitation energies. This is
essentially obtained as the incoherent sum of the effect generated by
the HF calculation and that obtained by the RPA.

The study of the excitation of the first two 1$^+$ excited states in
\lead \/ indicates that the lower energy state is more sensitive to the
tensor force than the other one. The presence of the tensor force
modifies energies and $B(M1)$-values, and slightly improves the
agreement with the experimental data \cite{shi08}, even though
the limitations of the RPA theory do not allow a description
of the fragmentation of the strength of the higher energy state. 

We have restricted our study to charge conserving nuclear excitations.
There are indications that the effects of the tensor-isospin force are
more relevant on the charge-exchange excitations
\cite{bai09a,bai09b,bai10}. The presence of the tensor forces seems to
be relevant also in the description of the reactions between heavy
nuclei \cite{iwa11}.

We think that the accuracy required today by self-consistent effective
theories requires the use of interactions containing tensor terms. 

\acknowledgments This work has been partially supported by the Spanish
Ministerio de Ciencia e Innovaci\'on under contracts
FPA2009-14091-C02-02 and ACI2009-1007 and by the Junta de
Andaluc\'{\i}a (FQM0220).


%
%
\clearpage
\newpage

\end{document}